\documentclass{sf2a-conf2023}
\usepackage{graphicx}
\usepackage{hyperref}
\usepackage[]{natbib}  
\usepackage{epstopdf}
\usepackage{bm}
\usepackage{amsmath}

\def\BibTeX{{\rm B\kern-.05em{\sc i\kern-.025em b}\kern-.08em
    T\kern-.1667em\lower.7ex\hbox{E}\kern-.125emX}}
\bibpunct{(}{)}{;}{a}{}{,}  %%%%%%%%%%%%%  A&A bibliography style
\newcommand{\ikd}{\mathrm{Im}\left[k_2^2(\omega)\right]}
\newcommand{\ep}{\epsilon_\mathrm{t}}
\newcommand{\qpo}{Q_\omega'}

\begin{document}

\TitreGlobal{SF2A 2023}

%%-----------------------------------------------------------------
%%      the top matter
%%

\title{Do nonlinear effects disrupt tidal dissipation \\ predictions in convective envelopes?} 

\runningtitle{nonlinear tidal dissipation}

\author{A. Astoul}
\address{Department of Applied Mathematics, School of Mathematics, University of Leeds, Leeds, LS2 9JT, UK
}

\author{A. J. Barker$^1$}

%% Keep this line, even if the page will be settled afterwards.
\setcounter{page}{237}

%%-----------------------------------------------------------------

\maketitle

%%-----------------------------------------------------------------
%%        The abstract
%% 
%%  Warning!  within the abstract:
%%  - do not use macros. 
%%  - do not use commands like: \cite, \citet, \citep ... etc.

\begin{abstract}
 Most prior works studying tidal interactions in tight star/planet or star/star binary systems have employed linear theory of a viscous fluid in a uniformly-rotating two-dimensional spherical shell. However, compact systems may have sufficiently large tidal amplitudes for nonlinear effects to be important. We compute tidal flows subject to nonlinear effects in a 3D, thin (solar-like) convective shell, spanning the entire frequency range of inertial waves. Tidal frequency-averaged dissipation predictions of linear theory with solid body rotation are approximately reproduced in our nonlinear simulations (though we find it to be reduced by a factor of a few), but we find significant differences, potentially by orders of magnitude, at a fixed tidal frequency corresponding to a specific two-body system at a given epoch. This is largely due to tidal generation of differential rotation (zonal flows) and their effects on the waves. 
 \end{abstract}

%% Insert the keywords (to appear in the ADS indexing)
%% Keywords must be separated by a comma

%%-----------------------------------------------------------------

\section{Introduction}
%%---------------------
Tidal interactions play an important role in driving spin-orbit evolution in close planetary and stellar binary systems.  A key mechanism in low-mass stars and giant planets is tidal dissipation of inertial waves (restored by the Coriolis acceleration) in their convective envelopes \citep[e.g.][]{OL2004,B2022}. For sufficiently compact systems, like for Hot Jupiter-like planets orbiting in a day or less around their host stars, the tidal flows are likely to be sensitive to nonlinear effects \cite[e.g.][]{O2014}.
We build upon the numerical model of \cite{AB2022} to simulate the nonlinear hydrodynamical dissipation of tidally-excited inertial waves in a 3D spherical shell containing an incompressible, viscous and adiabatically-stratified fluid. We aim to investigate the robustness of linear estimates of tidal dissipation derived for solid body rotation. For this purpose, we compute the imaginary part of the Love number $\ikd$ and the associated modified tidal quality factor  $\qpo=3/\left|2\,\mathrm{Im}\left[k^2_2(\omega)\right]\right|$ over the entire range of frequencies allowed for inertial waves. Here, we focus on large tidal forcing amplitudes relevant for close giant gaseous planets, and a thin (solar-like) convective shell (representing either the star or the outer envelope of a giant planet) with fractional inner radius $\alpha=0.7$. 
%%-------------------------
\section{Tidal dissipation with nonlinear effects} 
\begin{figure}[t!]
 \centering
 \includegraphics[trim=0cm 0.3cm 0cm 0.1cm,clip,width=0.9\textwidth]{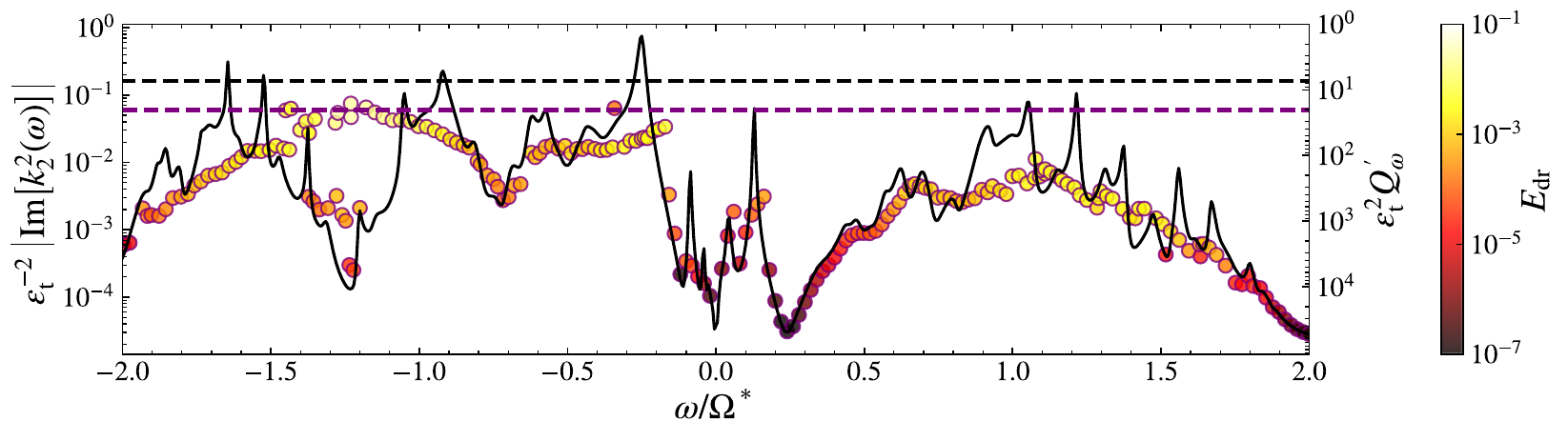}      
  \caption{Imaginary part of the Love number $\ikd$, or equivalently, modified tidal quality factor $\qpo$, both re-scaled (so system-independent), versus the Doppler-shifted frequency $\omega$ in the rotating frame with mean rotation $\Omega^*$. Parameters are $\alpha=0.7$, $C_\mathrm{t}=5\cdot10^{-2}$, and Ekman number $\mathrm{E}=\nu/(\Omega R^2)=10^{-5}$ (with $R$ the radius and $\Omega$ the initial rotation). Linear predictions are shown in black (the black dashed line is $\Lambda_\mathrm{l}$), while nonlinear ones are shown in reddish colour ($E_\mathrm{dr}$ quantifies the strength of zonal flows), while the frequency-averaged nonlinear $\Lambda$ is shown by the purple dashed line.}
  \label{fig1}
\end{figure}

Using the pseudo-spectral code MagIC (\url{https://magic-sph.github.io/}) amended to study tidally-excited inertial waves (of velocity $\bm u$ and forcing frequency $\omega$) in convective envelopes, we compute the volume-averaged ($\langle\cdot\rangle$) dimensionless tidal dissipation  $D_\nu(\omega)=-\left<\nu\bm u \cdot\Delta \bm u\right>$ with (atomic or turbulent) viscosity $\nu$ \citep[see][for more details]{AB2022,AB2023}. In our simulations, the emergence of cylindrical differential rotation triggered by nonlinear self-interactions of inertial waves causes large departures from linear predictions with solid body rotation. This is particularly true for a thin shell with high tidal forcing amplitudes and a low viscosity, as shown in Fig. \ref{fig1}. We display the frequency-dependent and frequency-averaged tidal dissipation: 
\begin{equation}
\left|\ikd\right|=\frac{6}{5}\frac{D_\nu(\omega)}{|\omega|}\frac{\ep^2}{C_\mathrm{t}^2},~~\mathrm{ and }~~\Lambda\equiv\int_{-\infty}^{+\infty}\ikd\,\frac{\mathrm{d}\omega}{\omega},
\end{equation}
with the tidal forcing amplitude $C_\mathrm{t}=\left(1+\mathrm{Re}\left[k^2_2\right]\right)(M_2/M_1)(R/a)^3$, and $\ep=\left(1+\mathrm{Re}\left[k^2_2\right]\right)\Omega/\sqrt{GM_1/R^3}$. 
For a homogeneous convective shell, $\Lambda_\mathrm{l}=16\pi\,\ep^2\alpha^5/[63(1-\alpha^5)]$ in linear theory \citep{O2013}.
\subsection{Tidal dissipation in HAT-P-65 b}
As an example, we calculate the tidal dissipation due to eccentricity tides in the Hot Jupiter HAT-P-65 b, which may have an eccentricity up to  $e\approx0.3$\footnote{From the \url{http://exoplanet.eu/} database, also used to calculate $C_\mathrm{t}$ and $\ep$.}. By assuming the planet's rotation is quasi-synchronised with its orbit, we have $\omega/\Omega^*=\pm1$ for the dominant tidal component. By  setting $\mathrm{Re}\left[k^2_2\right]=0.6$ \citep{DL2022}, we estimate that the dominant amplitude for eccentric tides is $7e\,C_\mathrm{t}/2\approx5\cdot10^{-2}$ \citep[][]{AB2023}, and $\ep\approx0.27$  
which means the planet is strongly deformed by its rotation. Then, using Fig.~\ref{fig1} (i.e.~assuming the planet has a convective shell with $\alpha=0.7$), the linear and nonlinear frequency-dependent tidal quality factors are, respectively,  
$\qpo\sim8\cdot10^2$ and $\qpo\sim3\cdot10^3$ for $\omega=1$. Therefore, taking into account nonlinear effects reduces tidal dissipation in HAT-P-65 b by almost a factor 4. For comparison, the linear frequency-averaged tidal quality factor, $Q'\sim10^2$, being dominated by the most dissipative resonant peaks, is more than 1 order of magnitude smaller than the frequency-dependent nonlinear value (hence the latter is less dissipative). 
\section{Conclusions}
%%--------------------
We have performed a nonlinear exploration of tidal dissipation estimates in convective envelopes representing low-mass stars or giant gaseous planets in close systems, over a large range of forcing frequencies \citep[different shell thicknesses, tidal amplitudes, and viscosities, are discussed in][]{AB2023}. Though nonlinear frequency-averaged tidal dissipation estimates are only moderately attenuated from prior linear predictions, nonlinear frequency-dependent rates can strongly differ from the linear ones (and from the frequency-averaged value), particularly when the shell is thin (large $\alpha$), the viscosity is low, and the tidal forcing is strong, as evidenced for our application to HAT-P-65 b. Differential rotation, especially the interactions between inertial waves and zonal flows, play a key role in smoothing and mitigating the highly frequency-dependent peaky tidal dissipation spectra (see Fig.~\ref{fig1}). Our results indicate that a more realistic treatment of tidal flows including nonlinear effects/differential rotation is necessary to model the spin/orbit evolution of a close two-body system.
\vspace{-5ex}
% Optional acknowledgements
% -------------------------
\begin{acknowledgements}
Funded by STFC grants ST/S000275/1 and ST/W000873/1, and by a Leverhulme Trust Early Career Fellowship to AA. 
\end{acknowledgements}

%%-----------------------------
%%   Bibliography
%%-----------------------------
%%
%% The reference list should contain all the references cited in the text, ordered alphabetically by surname (with
%% initials following). If there are several references to the same first author, they should be entered according
%% to the following scheme:
%% 1. One author: chronologically
%% 2. Author, one co-author: alphabetically by co-author, then chronologically
%% 3. Author, two or more co-authors: chronologically.
%%
%% Please note that for papers that have more than five authors, only the first three should be given, followed
%% by "et al."
%%
%% The format for references is the one adopted by A&A (see the example below).
%%
%% To set the reference list in the proper A&A format, we encourage you to use BibTEX and the natbib
%% package instead of the standard 'thebibliography' environment.
%%
%% The following lines are required when using BibTEX (strongly encouraged!):
\bibliographystyle{aa}  % A&A bibliography style file (aa.bst)
\bibliography{astoul} % your references in file: Yourfile.bib

\end{document}